\documentclass[a4paper,12pt]{article}
\usepackage{amsfonts,amssymb, xcolor,pdfpages,graphicx}

\title{Multiplication of Distributions and Nonperturbative Calculations of Transition Probabilities}
 
\author{J. Aragona, IME, USP, Sao Paulo, SP, Brazil,\\P. Catuogno, IME, UNICAMP, Campinas, SP, Brazil,\\J.F. Colombeau, Institut Fourier, Grenoble, France,\\ and IME, UNICAMP, Campinas, SP, Brazil,\\ S.O. Juriaans, IME, USP, Sao Paulo, SP, Brazil,\\ C. Olivera, IME, UNICAMP, Campinas, SP, Brazil. }

\date{}
\begin{document}
\maketitle

\abstract{In a mathematical context in which one can multiply distributions the "`formal"' nonperturbative canonical Hamiltonian formalism in Quantum Field Theory makes sense mathematically, which can be understood a priori from the fact the so called "`infinite quantities"' make sense unambiguously (but are not classical real numbers). The perturbation series does not make sense. A novelty appears when one starts to compute the transition probabilities. The transition probabilities have to be computed in a nonperturbative way which, at least in simplified mathematical examples (even  those looking like nonrenormalizable series), gives real values between 0 and 1 capable to represent probabilities. However these calculations should be done numerically and we have only been able  to compute  simplified mathematical examples due to the fact these calculations appear very demanding in the physically significant situation with an infinite dimensional Fock space and the QFT operators.}\\ \\

\textbf{1. Mathematical background}

In 1954 L. Schwartz published a note
\textit{Impossibility of the multiplication of distributions} \cite{Schwartznote}
 and he claimed \cite{Schwartzlivre}
\textit{Multiplication of distributions is impossible in any mathematical context, possibly different of distribution theory.}\\
 This means therefore \textit{for ever}.
The viewpoint of mathematical physicists was then that multiplications of distributions in physics emerged from an erroneous mathematical formulation of physics to be replaced by a correct mathematical formulation to be discovered: this was at the origin of the development of various "`axiomatic"' field theories in which physics was summed up in a list of "`axioms"' and the mathematical difficulty was shifted to the construction of mathematical objects satisfying the axioms.\\ 

Thirty years  later in 1983 L. Schwartz presented to the french Academy of Sciences the note \cite{Colombeaunote}
\textit {A General Multiplication of Distributions} 
whose title is  exactly the converse of the one of his 1954 note \cite{Schwartznote}.
 The detailed proofs of the note were published in form of two books
  \cite{Colombeaulivre1}, 1984 and
\cite{Colombeaulivre2} 1985.
	A clarification is needed.\\ \\

\textbf{2. A mathematical context in which one can multiply distributions.}

This clarification is obvious from the two formulas (where H denotes the  Heaviside function $H(x)=0$ if $x<0, H(x)=1 $ if $x>0, H(0)$ unspecified):
\begin{equation}\int(H^2-H)(x)\psi(x)dx=0 \ \ \forall \psi\in \mathcal{C}_c^\infty,\end{equation}
\begin{equation}\int(H^2-H)(x)H'(x)dx=[\frac{H^3}{3}-\frac{H^2}{2}]_{-\infty}^{+\infty}=-\frac{1}{6},\end{equation}
which both should hold in a context in which one can multiply distributions (so that $(H^2-H)H'$ makes sense), see\cite{Varna}.\\  

First we give a coarse incorrect proof that these two formulas are contradictory: formula (1) implies $$  H^2-H=0.$$
Formula (2) implies $$  H^2-H\not=0,$$ hence a contradiction which proves impossibility of existence of a mathematical context in which one could multiply distributions. \textit{Where is the  mistake?}\\

The mistake is that we assume we are in a new unknown mathematical context and we ignore if in this context 
\begin{equation}(\int F(x) \psi(x) dx=0 \ \forall \psi\in \mathcal{C}_c^\infty)\Rightarrow F=0\end{equation}
implicitly used by habit to claim $H^2-H=0$ (in the 1954 Schwartz proof this follows from a property imposed to any reasonable multiplication of distributions). The correct conclusion is:\\
 \\
\textit{The familiar implication (3)
ceases to be valid in a context in which one can multiply distributions.}\\
 \\
This causes no trouble for a differential calculus similar to the classical one: in 1982 one of the authors  published such a context \cite{ColombeauJMAA}, later developed in the books and surveys \cite{Colombeaulivre1,Colombeaulivre2,Colombeaulecturenotes,ColombeauBAMS,AraBia,Bialivre,GKOS,AraFerJu,SteinVi,Gsponerseul,Vi} among other.
The first novelty is simply the need to introduce a new notation:
$$ F\approx G \hbox{ iff by definition } \int (F-G)(x)\psi(x)dx=0 \ \forall \psi\in \mathcal{C}_c^\infty$$ 
which is some kind of weak relation that generalizes exactly the equality of distributions. We call this relation "`association"'. It has the properties and the peculiarities of distributions: 
$$F\approx G \Rightarrow F'\approx G' \ \hbox{ but } \not\Rightarrow FK\approx GK$$
 if $K$ is another generalized function.
As a basic example $$H^2\approx H \hbox{ and } H^2\not=H.$$
 As another example one can define easily various natural positive square roots of the Dirac distribution $\delta$ and one has 
 $$\sqrt{\delta}\approx 0 \hbox{ and } \delta=(\sqrt{\delta})^2 \not\approx 0$$ showing again incoherence between $\approx$ and nonlinear operations.\\  

\textit{In short one has defined a basically new concept of equality, more restrictive than the classical one. Then the natural extension of the classical concept of equality is given the name of "`association"' and the notation $\approx$. In the applications of this new theory one has to play with = and $\approx$ at their right place.} Basically this is very simple.\\


Let $U$ and $V$ be two distributions from physics that cannot be multiplied within the distributions and for which the product $UV$ is  usually considered as ambiguous. What does this theory give? A priori nothing more: if $F_i\approx U$ and $G_j\approx V$ (there are many 
$F_i$ and $G_j$) then the various products $F_i G_j$ take different values: one recovers the ambiguity.  But a novelty:\\  \\
\textit{ In this theory one can state more precisely than usual, on physical ground, the equations: this can give some selection on $F_i$ and $G_j$ which can resolve the ambiguity: then one can obtain new formulas to be compared with experimental results}.\\

This has been done for the equations modeling strong collisions used to design the armor of battle tanks: \cite{Colombeaulecturenotes}. From a differential statement of Hooke's law used by engineers these equations are in nonconservative form which  in the case of shock waves gives rise to products of distributions not defined within the distributions. A suitable formulation of the equations in this new context, motivated on physical ground, consisting in stating some equations of the system with $=$ and the other ones with $\approx$ permits to resolve the ambiguity. Then one obtains well defined formulas ruling the shock waves for this system, which have been checked to be in agreement with the experimental observations.  \\ \\


\textbf{3. A brief recall on the calculations in the canonical Hamiltonian formalism for a neutral massive self-interacting boson field.}

The Hilbert space of states called Fock space is the Hilbertian direct sum $$\mathcal{F}=\oplus_{n=0}^\infty L^2_s((R^3)^n),$$ 
where the subscript $s$ means symmetric in the $n$ arguments in $R^3$. 
The basic linear operators used in $\mathcal{F}$ are the creation and annihilation operators: if $\psi\in L^2(R^3)$ the creation operator $a^+(\psi)$ is the unbounded operator on $\mathcal{F}$ defined by

$$(f_n)\longmapsto (0,f_0\psi,\dots,\sqrt{n}Sym(f_{n-1}\otimes\psi),\dots)$$
where $Sym$ is the symmetrization operator. The annihilation operator  $a^-(\psi)$ is given by the formula 
$$(f_n)\longmapsto (\int f_1(\lambda)\psi(\lambda)d\lambda, \dots,\sqrt{n+1}\int f_{n+1}(\lambda)\psi(\lambda)d\lambda,\dots).$$


The free field operator is given by the formula:\\ 
\\ 
$\Phi_0(x,t):=a^+(k\mapsto 2^{-\frac{1}{2}}(2\pi)^{-\frac{3}{2}}(k^0)^{-\frac{1}{2}} e^{ik^0t}e^{-ikx})+$ $$a^-(k\mapsto 2^{-\frac{1}{2}}(2\pi)^{-\frac{3}{2}}(k^0)^{-\frac{1}{2}} e^{-ik^0t}e^{+ikx})$$ 
where  $k^0=\sqrt{k^2+m^2}$.\\

If $\Phi(x,t)$ denotes the (unknown) interacting field operator the total Hamiltonian operator is given by the formula  
\begin{equation}H(t):=\int[\frac{1}{2}(\partial_t\Phi)^2+\frac{1}{2}\sum_{i=1}^3(\partial_i \Phi)^2+\frac{m^2}{2}\Phi^2+\frac{g}{N+1}\Phi^{N+1}](x,t) dx.\end{equation}
Another related important operator denoted $H_0(t)$ is obtained by inserting the free field operator $\Phi_0$ into (4) in place of the interacting field operator $\Phi$.\\


The Hamiltonian formalism consists in formal calculations on the free field operator that produce the interacting field operator and the scattering operator, then transition probabilities that are compared with experimental results.
We consider two time values $\tau<t$.  
The interacting field operator is given by the formula
$$\Phi(x,t):=e^{i(t-\tau)H_0(\tau)}\Phi_0(x,\tau)e^{-i(t-\tau)H_0(\tau)}.$$
Formal calculations give that it satisfies the "` interacting field equation"'
$$\partial_t^2\Phi(x,t)=\sum_{i=1}^3\partial^2_i\Phi(x,t)-m^2\Phi(x,t)-g\Phi^{N}(x,t).$$
The scattering operator is given by the formula:
\begin{equation}S_{\tau}(t) :=e^{i(t-\tau)P_0}e^{-i(t-\tau)H_0(\tau)},\end{equation} where  $P_0$ is the energy operator
 defined by the formula: if $F=(f_n)\in \mathcal{F}$

$$ (P_0) F=(0,k\mapsto k^0f_1(k),\dots,(k_1,\dots,k_n)\mapsto (k_1^0+\dots+k_n^0)f_n(k_1,\dots,k_n),\dots).$$


Formal calculations give that the scattering operator permits to obtain the interacting field operator $\Phi(x,t)$ from the free field operator at same time $\Phi_0(x,t)$ through the formula
$$ \Phi(x,t)=(S_{\tau}(t))^{-1}\Phi_0(x,t)S_{\tau}(t).$$
Formal calculations give that the scattering operator satisfies the ODE
$$ \partial_tS_{\tau}(t)=-i\frac{g}{N+1}\int\Phi_0(x,t)^{N+1}dx \ S_\tau(t), \ \ S_\tau(\tau)=id.$$
The transition probabilities are given by the formula
\begin{equation}|<F_1,S_\tau(t)F_2>|, \ F_1,F_2\in \mathcal{F}.\end{equation}
These formal calculations are recalled in detail in \cite{Colombeaulivre1,ColombeauarXiv}.\\ \\


\textbf{4.  Interpretation of QFT in this mathematical context.}

In order to interpret the canonical Hamiltonian formalism in the context of nonlinear generalized functions we need  an heuristic understanding of the new context:\\
a real number, a distribution, a nonlinear generalized function,
are all merely:\\
(a regularization) modulo an imprecision (the imprecision appears mathematically as a mathematical quotient). In more detail:\\
\\
$\bullet$ A real number is a Cauchy sequence $(x_n)$  of rational numbers modulo the set of  \{all sequences that tend to 0\}.\\
$\bullet$ A distribution is a sequence $(f_n)$  of $\mathcal{C}^\infty$ functions having the property that  $ \forall \psi \int f_n\psi dx\rightarrow$ a limit  modulo the set of $\{\hbox{all sequences } (g_n) \hbox{  such that } \forall \psi \int g_n\psi dx\rightarrow 0\}$.\\
$\bullet$ A nonlinear generalized function is a certain sequence $(f_n)$ of $\mathcal{C}^\infty$ functions modulo the set of $\{$all sequences that tend to 0 in a certain sense$\}$.\\

This interpretation of real numbers is well known. This interpretation of distributions is given in \cite{Antosik}. We observe that this is no more than the repetition of a standard process in mathematics: \textit{regularization of a "`mysterious"' new object} to define it by means of known objects and then \textit{imprecision} to diminish the number of new objects so constructed by regularization.\\

How to interpret QFT with these nonlinear generalized functions?\\


The free field operator $\Phi_0(x,t)$ is a  distribution in $x$ for each fixed $t$. To produce the needed family of $\mathcal{C}^\infty$ functions that should represent it  we use a regularization by convolution. We set 
\begin{equation}\Phi_0(x,t,\epsilon)=(\Phi_0(.,t)*\phi_\epsilon)(x), \ \ \phi_\epsilon(x)=\frac{1}{\epsilon^3}\phi(\frac{x}{\epsilon}),\ \int\phi dx=1.\end{equation}
This depends on a choice of $\phi$ and one should check that the final result of the theory should be independent of this choice.\\
   \\
The calculations in section 3 make sense from \\ 
   
\textbf{Theorem 1}. \textit{The symmetric operators that occur (such as the total Hamiltonian $H_0(\tau)$) admit a  self-adjoint extension.}\\ \\
  A proof is in \cite{Colombeaulivre1, Gsponer}.	
This permits to define the imaginary exponentials such as $e^{-i(t-\tau)H_0(\tau)}$.
In particular one defines the scattering operator (5), then the transition probabilities (6), that both are equivalence classes, in the new mathematical context, of representatives depending on the choice of a function $\phi$ in $\phi_\epsilon$ in (7).\\

Let $F_1, F_2\in \mathcal{F},$ of norms=1.
\textit{What is the transition probability $|<F_1,S_\tau(t)F_2>|$}?\\
\\
The generalized operator $S_\tau(t)$ is the equivalence class of the family $\{S_\tau(t,\epsilon)\}$ that depends on $\epsilon$ and, for each $\epsilon$, $|<F_1,S_\tau(t)(\epsilon)F_2>|$ is a real number between 0 and 1. Usually it oscillates infinitely between 0 and 1 when $\epsilon\rightarrow 0$ like $|cos\frac{1}{\epsilon}|$: this is due to the presence of "`infinities"' such as $\frac{1}{\epsilon}$ when $\epsilon\rightarrow 0$.\\

In the theory developed in this paper the  infinities are transformed into infinite oscillations because we are in a nonperturbative formulation and the infinities appear inside imaginary exponentials such as $exp(\frac{i}{\epsilon})$.\\
 \\
\textit{How to interpret these infinite oscillations?} since we wish  a well defined real number between 0 and 1 to be interpreted as a probability. One has to
obtain the answer from  the interpretation of nonlinear generalized functions in physics.\\ \\


\textbf{5. Calculations of an infinite number of oscillations.}

In classical mechanics an experiment is represented by a value of $\epsilon>0$ very small, denoted $\epsilon_0$, that can only slightly change from an experiment to another due to small differences in experimental conditions: one can eliminate the influence of parasites (except in situations of turbulence). Since, in the examples from  classical mechanics, the result, depending on $\epsilon$, has a limit when $\epsilon\rightarrow 0$ this limit is a very good approximation of the result obtained from the value $\epsilon_0$.\\
\ \\
Here, in the calculations of QFT,  we have no limit but infinite oscillations between 0 and 1. The origin of these oscillations can be attributed to the influence of the  void on the interaction itself (fictuitous particle-antiparticle pairs) that we cannot eliminate and is aleatory. Therefore different experiments can give very different results. Since the final experimental result is an average (a probability) then the same should be done at the level of the theory.\\
\  \\
This suggests that the result to be compared with experiments should be
 \begin{equation}\frac{1}{N}\sum_{i=1}^N|<F_1,S_\tau(t,\epsilon_i)F_2>|, \ N \hbox{ very large  },  \epsilon_i>0 \hbox { very small, at random }.\end{equation}
Does such a mean value exist? One expects YES.
This expectation is supported by two kinds of arguments: numerical calculations in very simplified cases and a theoretical proof also in a simplified case.   \\

Numerical calculations, from the definition of mean value, in Hilbert spaces of dimension $n=$ 2 and 3,  can be done on a PC very easily:  it suffices to compute a large number of times ($j=1,\dots,N,  N $ very large) the quantity
$|<F_1,exp(iH(\epsilon_j))F_2>|$ if $H(\epsilon)$ is a hermitian  symmetric $n\times n$ matrix whose coefficients are functions of $\epsilon$ that can tend to $\infty$ when $\epsilon\rightarrow 0.$\\
\  \\
A theoretical result is as follows:\\ \\
\textbf{Theorem 2.}  \textit{ In a Hilbert space of finite dimension $n$ let $H$ be a $n\times n$ symmetric matrix whose coefficients are "`reasonable"' functions of $\epsilon$ that can tend to $\infty$ when $\epsilon\rightarrow 0$ then $|<F_1,exp(iH(\epsilon))F_2>|$ has a mean value when $\epsilon\rightarrow 0$.}\\ \\
The word reasonable means: polynomials and exponentials in $\frac{1}{\epsilon}$, their quotients with nonzero denominator, etc.\\

Our proof is long and difficult. It uses results from the theory of almost periodic functions \cite{Besi,Bohr}.  We have not been able to extend the proof to the Fock space and the operators of QFT presumably because the matter  becomes very complicated. The case of QFT looks exactly similar at a heuristic level.\\

Here are a few numerical tests that can be reproduced at once on any PC.\\

Example: mean value of $|<u,exp(-iH)v>|, \ u=(\frac{\sqrt{2}}{2},\frac{\sqrt{2}}{2}), v=(\frac{\sqrt{2}}{2},-\frac{\sqrt{2}}{2}),$
$H=$ first line $(\frac{a}{\epsilon^{n_1}+\epsilon^{2n_1}}, \frac{c}{\epsilon-2\epsilon^2})$, second line $(\frac{c}{\epsilon-2\epsilon^2},\frac{b}{\epsilon^{n_2}})$\\
 \\
$a=1,b=1.3,c=0.4,n_1=1,n_2=1,\epsilon<10^{-7}$ at random. We obtained\\
\  \\
$N=10^5$ \ \ 0.223 ten times, 0.224 two times\\
$N=10^7$ \ \ 0.2235 ten times, 0.2236 two times\\
$N=10^8$ \ \ 0.22353 six times, 0.22352 five times, 0.22351 one time\\
 \\ Different values: $a=1,b=1,c=2,n_1=0.5,n_2=1,\epsilon<10^{-7}$ at random. We obtained\\
 \\
$N=10^5$ \ \ 0.1539 one time, 0.154 ten times, 0.1551 one time\\
$N=10^6$ \ \ 0.1544 eight times, 0.1543 four times.\\

Example: mean value of $\frac{1}{\sqrt{6}}|exp(\frac{i}{\epsilon})-2 exp(\frac{i}{\epsilon^2})+(1+\epsilon)exp(\frac{i}{\sqrt{\epsilon}})|$ with $\epsilon<10^{-10}$ at random\\
 \\
$N=10^4$: \ 0.9313, \ 0.9188, \ 0.9212 \ \ =0.9 for sure\\
$N=10^5$: \ 0.9257, \ 0.9258, \ 0.9243  \  \ =0.92 for sure\\
$N=10^6$: \ 0.9249, \ 0.9248, \ 0.9247  \  \ =0.924 for sure?: more trials needed\\
$N=10^7$: \ 0.9249, \ 0.9251, \ 0.9250\\
$N=10^8$: \ 0.9251, \ 0.9251, \ 0.9250  \ \ =0.925 for sure\\
$N=10^9$: \ 0.9251, \  0.9251, \ 0.9251 \  \ =0.9251  for sure?: more trials needed\\
 \\

The calculations from  (8) show that one needs large values of $N$ to reach the mean value even in such  very simple cases. This method is not efficient enough for large dimension a fortiori for the Fock space. This mean value appears very robust presumably independent on the choice of 
$\phi_\epsilon$ in (7).\\ \\

\textbf{6. Conclusion.}

What to do now?.\\
 - try to finish the proof to get closer  to QFT.\\
 \\
 -find a significantly better numerical method that could apply in large dimensions, then in QFT.\\
 \\
 -try to compare the result from the mean value with Renormalization in the case of renormalizable theories.\\
 \\
Does there exist a mathematically similar physical theory in a finite dimensional Hilbert space that could serve for a validation of the method of mean values presented in this paper by comparison theory-experiments?\\
  \\
Our conclusion is that the mathematical interpretation of QFT in this paper is likely a possible nonperturbative QFT. However it is  unfinished and there remains great difficulties that appear of a technical nature.


\textbf{Acknowledgement.} \textit{One of the authors (JFC) is very much indebted to Raymond Stora for help and discussions in the years 1976-1977, which were at the origin of his  work on multiplication of distributions and of the contents of this paper.
P. Catuogno and C. Olivera  are partially supported  FAPESP by the grant 2015/07278-0.  C. Olivera  is partially supported  FAPESP by the grant 2017/17670-0.
}

 J. Aragona, aragona@ime.usp.br,
 P. Catuogno, pedrojc@ime.unicamp.br,
 J.F. Colombeau jf.colombeau@wanadoo.fr,
 S.O. Juriaans, ostanley16@yahoo.com.br, 
C. Olivera, colivera@ime.unicamp.br.

\end{document}